\documentstyle[aps,prl,twocolumn]{revtex}

\newcommand\ee{\end{equation}}
\newcommand\be{\begin{equation}}
\newcommand\eea{\end{eqnarray}}
\newcommand\bea{\begin{eqnarray}}

\newcommand\GeV{\,\mbox{GeV}}

\newcommand\mpl{m_{Pl}}

\newcommand\lsim{\mathrel{\rlap{\lower4pt\hbox{\hskip1pt$\sim$}}
    \raise1pt\hbox{$<$}}}
\newcommand\gsim{\mathrel{\rlap{\lower4pt\hbox{\hskip1pt$\sim$}}
    \raise1pt\hbox{$>$}}}

\begin{document}

\par
\begingroup
\twocolumn[%
\vspace*{-3ex}
\hspace*{\fill}{LANCASTER-TH/9612}\hspace*{3.5em}\\
\hspace*{\fill}{hep-ph/9606387}\hspace*{3.5em}
\vskip 10pt
{\large\bf\centering\ignorespaces
What would we learn by detecting a gravitational wave signal in the
cosmic microwave background anisotropy?
\vskip2.5pt}
{\dimen0=-\prevdepth \advance\dimen0 by23pt
\nointerlineskip \rm\centering
\vrule height\dimen0 width0pt\relax\ignorespaces
David H.~Lyth 
\par}
{\small\it\centering\ignorespaces
School of Physics and Chemistry, Lancaster University,
Lancaster LA1 4YB,~~~U.~K. \\
\par}
{\small\rm\centering(\ignorespaces June 1996\unskip)\par}
\par
\bgroup
\leftskip=0.10753\textwidth \rightskip\leftskip
\dimen0=-\prevdepth \advance\dimen0 by17.5pt \nointerlineskip
\small\vrule width 0pt height\dimen0 \relax
Inflation generates gravitational waves, which may be observable in the
low multipoles of the cosmic microwave background (cmb) anisotropy
but only if the inflaton field variation
is at least of order the Planck scale. Such a large variation
would imply that 
the model of inflation cannot be part
of an ordinary extension of the standard model, and 
combined with the detection of the waves it
would also suggest 
that the inflaton field cannot be one of the superstring 
moduli.  Another implication of observable gravitational waves would be
a potential $V^{1/4}=2$ to
$4\times 10^{16}\GeV$, which is orders of magnitude bigger than is 
expected on the basis of particle theory. It might emerge in a hybrid 
inflation model where most of the energy density comes from the 
Higgs sector of a GUT, but only if both the vacuum expectation values
{\em and the masses
} of the Higgs fields are of this order.
\par\egroup
\vskip2pc]
\thispagestyle{plain}
\endgroup


\section{Introduction}

Inflation generates a density perturbation and gravitational waves.
The density perturbation is thought to be responsible for
large scale structure and, together with a possible gravitational 
wave contribution, for 
the cosmic microwave background (cmb) anisotropy.
It is well known that the detection of a gravitational wave
contribution to the cmb anisotropy would 
immediately determine the value $V$ and slope $V'$ of the potential
while relevant scales are leaving the horizon during inflation
\cite{LL2}, with an eventual
measurement of the spectral index of the density perturbation 
fixing $V''$ \cite{LL1,LL2} and 
additional data providing limited additional
information about the shape of $V$ \cite{recon}.
Here I point out that a detection would also tell one
that the inflaton field variation during inflation is at least
of order the Planck scale, and go on to
discuss the 
theoretical implications of both this result and the value of $V$.

If $\delta_H^2$ is the spectrum of the curvature perturbation 
associated with the density perturbation,
and ${\cal P}_g$ is the spectrum of the gravitational 
waves (as defined for example in \cite{LL2}),
it is convenient to consider the ratio
$r(k)=0.139 {\cal P}_g/\delta_H^2$.
The spectra are in general scale dependent, and $r(k)$ 
has been normalized
so that it in an analytic approximation \cite{starobinsky}
it gives the 
ratio of the two contributions
to the mean-square quadrupole of the 
cmb  anisotropy seen by a randomly placed observer. For higher multipoles 
the corresponding ratio is roughly constant in the range 
$1<l\lesssim 100$, but then it falls off sharply so that it will be 
detected if at all in the above range.

The standard slow-roll paradigm of inflation \cite{foot1}
predicts
\cite{adpred,lyth85,rubakov,starobinsky} 
\bea
\delta_H^2(k) &= & \frac1{75\pi^2\mpl^6} \frac{V^3}{V'^2} \\
r(k) &=& 6.9 \mpl^2(V'/V)^2
\eea
where $\mpl=(8\pi G)^{-1/2}=2.4\times 10^{18}\GeV$ is the Planck scale.
The right hand sides are
evaluated when $k=aH$ where $k/a$ is the wavenumber,
$a$ is the scale factor and $H=\dot a/a$.

In an interval $\Delta \ln k\sim 1$,
the fractional changes in $\delta_H^2$ and ${\cal P}_g$
are predicted to be $\ll 1$.
Since the 
$l$th multipole of the cmb anisotropy corresponds to a scale
$k^{-1} \simeq 2/(H_0 l) $ the relevant range $1<l\lesssim 100$
corresponds to only 
$\Delta \ln k\simeq 4.6$ so $r(k)$ will have a roughly constant value 
which from now on will be denoted simply by $r$.
Ignoring any variation one can show \cite{turner}
that because of cosmic variance a value $r>.07$ is necessary
in in order
to have a better than even 
chance of eventually detecting the gravitational wave contribution, and
approximately the same result should hold for the average even if there 
is some variation.

At present observation provides only a weak upper bound on $r$, which 
has not been quantified properly but is  something like
$r\lesssim 1$ \cite{andrewpers}. The COBE observations give a good 
normalization, \cite{delh} $\delta_H\simeq 1.9 (1+r)^{1/2}
\times 10^{-5}$, and using it one finds \cite{foot2}
\be
V^{1/4}\simeq (r/.07)^{1/4} \times 1.8\times 10^{16}\GeV
\label{vnorm}
\ee
Thus a detection of $r$ would give a value
$V^{1/4}=2$ to $4\times 10^{16}\GeV$. 

The slow-roll paradigm also gives
\be
\frac1\mpl \left|\frac{ d\phi}{d N} \right|= \mpl\left|\frac{V'}{V} \right|
=\left( \frac r{6.9} \right)^\frac12
\ee
where $d\phi$ is the change in the inflaton field 
in $dN=H dt\simeq d\ln a$ Hubble times.
While the scales corresponding to $1<l\lesssim 100$
are leaving the horizon $\Delta N
\simeq 4.6$, so the corresponding 
field variation is
\be
\Delta\phi/\mpl\simeq 4.6(r/6.9)^{1/2} =0.46(r/.07)^{1/2}
\label{deltaphi}
\ee
We see that a detectable $r$ requires $\Delta\phi\gtrsim 0.5\mpl$.
This is a minimum estimate for the total field variation, because
inflation continues afterwards for some number $N$ 
of $e$-folds.
The standard estimate \cite{LL2}
is $N\simeq 50$, but with late reheating and a single epoch of 
thermal inflation \cite{thermal1,thermal} $N\simeq 25$.
In either case it is clear that $r(k)$ can
increase significantly on smaller scales, making the 
total field variation much bigger than the estimate (\ref{deltaphi}).
In fact, there is a whole class of models where the increase is 
typically so 
strong that  a detectable $r$ requires $\Delta\phi\gg\mpl$.
These are the models where
the inflaton field is near a maximum of the 
potential.\footnote
{If the potential is 
$V\simeq V_0-\frac12 m^2 \phi^2$ one has 
$(\Delta\phi/\mpl)^2\simeq V_0/(\mpl^2 m^2) =2/(1-n) \gg 1$
(where $1-n$ is the spectral index), but
$r\simeq (6.9/2)(1-n)e^{-(1-n)N}<.051(25/N)$
which is undetectable. 
If $V\simeq V_0[1-(\phi/M)^p]$, with 
$p>2$ and
$M\gtrsim \mpl$, one has 
$\Delta\phi\simeq M\gtrsim \mpl$
but
$r=6.9p^2(M/\mpl)^\frac{2p}{p-2}[Np(p-2)]^{-\frac{2p-2}{p-2}}$
which is detectable only if $M$ is very big ($M>6.3\mpl$ if $p=3$
and $M>8.4\mpl$ if $p=4$). Similar results hold if 
$V$ is a mixture of terms, say quadratic
at small $\phi$ and quartic at larger $\phi$, 
provided that all terms have the same sign.}

Now let us consider inflation model-building in the light of all this.
In the earliest models \cite{new}
the inflaton field  is rolling towards a vacuum expectation value
(vev) $\langle \phi\rangle\ll\mpl$, 
making $\Delta\phi\ll\mpl$ and $r$ negligible.
These models were at best unattractive because the inflaton field had to 
be very weakly coupled,
so `primordial' models were suggested, where the vev 
and $\Delta\phi$ are of order $\mpl$ \cite{primordial}
or bigger \cite{nontherm}; these still give 
negligible $r$ because inflation takes place near a maximum of the 
potential.
Then power-law
potentials $V\propto \phi^p$ were considered \cite{chaotic},
 where the field during inflation is rolling towards the 
origin with a value and a variation of order $10\mpl$,
giving a detectable $r$.
Finally
(confining ourselves to the case of Einstein gravity) 
`hybrid' inflation 
has been
proposed \cite{l90}, where the inflaton field 
is accompanied by another field responsible for most
of the potential energy, inflation ending when it is 
destabilized.
Like the earliest models, typical hybrid inflation 
models have \cite{CLLSW}  $\Delta\phi\ll\mpl$ 
(and therefore negligible $r$)
but unlike them they need not involve very small couplings
\cite{foot3}.

Should we care whether the field variation is big or small, when 
building a model of inflation? 
In the context of global supersymmetry (or no supersymmetry)
the answer would be no, because $\mpl$ makes no appearance
in the field theory.
However, according 
to present ideas, the extension of the standard model chosen 
by nature is likely to involve supergravity. In that context, one
expects the potential to have an infinite power-series expansion 
in each field,
\be
V=V_0+\frac12 m^2\phi^2 + \lambda\phi^4 + \lambda' \mpl^{-2}\phi^6
+\lambda'' \mpl^{-4}\phi^8 +\cdots
\label{power}
\ee
(For simplicity I am supposing that odd powers are excluded by a 
symmetry.) Ordinary field theory corresponds to a truncation at low
order, which is justified if all fields are small.
This is indeed the case for the usual applications
of field theory, 
involving the standard model, its 
minimal supersymmetric extension and more ambitious extensions
invoking such things as neutrino masses, 
Peccei-Quinn symmetry or a GUT.

So the answer to the question is that we should care very much.
Small-field models, which in practice seems to mean
hybrid inflation models, 
are under relatively good control; it will be enough to keep 
one or two dominant, low-order terms in 
expansion (\ref{power}) of $V$
(with perhaps quantum corrections
\cite{new,qaisar})
and one can hope to further restrict 
$V$ by
requiring that the fields relevant for inflation already appear
in an extension of the standard model designed for some other 
purpose \cite{lisa}.  

If a gravitational wave effect is detected in the cmb anisotropy,
we shall need a model of inflation
in which the inflaton field is of order $\mpl$ or bigger.
For a generic field one has no idea what to expect in this regime.
The only exception is for the superstring moduli, 
where superstring theory provides some guidance. 
The moduli potential looks \cite{bingall,paul}
as if it might be marginally capable of supporting inflation, 
in that the expected values of $\mpl^2(V'/V)^2$ and $\mpl^2V''/V$ 
at a generic point are of order 1 so that there could be an exceptional 
region in the moduli space
where these quantities are both small. 
Investigations using specific models \cite{bingall,natural2,macorra}
have actually concluded that viable inflation does not occur,
but even if it does it 
will probably not give a detectable $r$.
The reason is that  one expects the size of the region in field space
where inflation can occur to be only of order $\mpl$, and
in order to motivate the initial condition by invoking eternal inflation
\cite{eternal}
one will probably want to start inflation near a maximum of the 
potential. As we saw earlier, the combination of these two requirements
will probably not give a detectable $r$.

The conclusion is that a
model of inflation 
giving a detectable 
$r$ will probably 
live in 
uncharted territory,
where there is as yet 
no theoretical guidance as to the form of the potential.
There is no particular reason to invoke 
the usually-considered forms $V\simeq A\pm B\phi^p $, 
though of course one 
should still test such forms against observation by measuring 
both $r$ and the 
spectral index of the density perturbation \cite{LL2}.

Finally let us return to the result that 
$V^{1/4}$ will have to be a few times $10^{16}\GeV$
if there is a detection.
It has pointed out by several 
authors \cite{mutated,scott,paul,graham,lisa} that
such a big inflationary
potential is difficult to understand on the basis of particle
theory, which might generically suggest a scale of order $(m\mpl)^{1/2}
$ or $(m\mpl^2)^{1/3}$ with $m\sim 10^2\GeV$. More particularly,
one does not {\em expect} such a potential to be generated by the
Higgs sector of a GUT, because this would give (at the maximum)
$V\sim m_h
^2 \langle \phi_h \rangle^2$ and although coupling constant unification
suggests vevs $\langle \phi_h \rangle\sim 10^{16}\GeV$ there is no reason for 
the masses $m_h$ to be so big \cite{mhiggs}. 
But in the face of a measured 
$V^{1/4}$ of this order one might set aside all prejudice,
and look at the viablity of a hybrid inflation model with a
GUT higgs as the non-inflaton field and a {\em large} inflaton field
variation. 

To summarize, the observation of a 
gravitational wave signal in the cmb anisotropy
would require a revision of current thinking about the likely form
of the inflationary potential, in respect of both the field variation 
and the height of the potential. Turning the viewpoint around,
it is fair to say that there is at present a considerable theoretical
prejudice against the likelyhood of  such an observation.

\underline{Acknowledgements}:
I am indebted to Ewan Stewart for discussions and correspondence about
supergravity and superstring phenomenology.
I acknowldege support from PPARC, and from
the European Commission under the Human Capital and Mobility
programme, contract no. CHRX-CT94-0423.


\begin{references}
\bibitem{foot1} The standard version of the slow-roll paradigm
supposes that there is an essentially unique inflationary trajectory,
with a potential $V(\phi)$
satisfying $\mpl^2(V'/V)^2\ll 1$ and $\mpl^2|V''/V|
\ll 1$, and an evolution given by
 $3H\dot\phi=-V'$. A more general paradigm 
\cite{misao} leads to the same
prediction for ${\cal P}_g$ but in general a bigger
prediction for $\delta_H$ which would strengthen our conclusions.
\bibitem{foot2} The factor $1+r$ has been ignored because the prediction for
$\delta_H$ has errors \cite{stly} of order $r$.
\bibitem{foot3}
To be precise, $\Delta\phi$ is small for any reasonable values
of the parameters in simple versions of hybrid inflation,
and can be made small without fine-tuning in more complicated
ones \cite{mutated,ournew}.
\end{references}
\end{document}